\documentclass[11pt]{article} 
\usepackage{hyperref} 
\pdfoutput=1 
 
\begin{document} 
 
\title{Summoning the wind: Hydrodynamic cooperation of forcibly ejected fungal spores} 
 
\author{Marcus Roper$^{1}$, Agnese Seminara$^{3}$, Ann Cobb$^{2}$,\\ Helene R. Dillard$^{2}$ and Anne Pringle$^{4}$ \\ 
\\ $^{1}$Dept. of Mathematics and Lawrence Berkeley Laboratory,\\University of California, Berkeley, CA 94720
\\ $^{2}$ Dept. of Plant Pathology, N.Y. State Agricultural Expt. Station\\ and Cornell University, Geneva, NY 14456
\\ $^{3}$ SEAS and $^{4}$ Dept. of Organismic and Evolutionary Biology,\\ Harvard University, Cambridge, MA 01238}

\maketitle 
 
\begin{abstract} 
The forcibly launched spores of the crop pathogen \emph{Sclerotinia sclerotiorum} must eject through many centimeters of nearly still air to reach the flowers of the plants that the fungus infects. Because of their microscopic size, individually ejected spores are quickly brought to rest by drag. In the accompanying fluid dynamics video we show experimental and numerical simulations that demonstrate how, by coordinating the nearly simultaneous ejection of hundreds of thousands of spores,\emph{Sclerotinia} and other species of apothecial fungus are able to sculpt a flow of air that carries spores across the boundary layer and around intervening obstacles. Many spores are sacrificed to create this flow of air. Although high speed imaging of spore launch in a wild isolate of the dung fungus \emph{Ascobolus} shows that the synchronization of spore ejections is self-organized, which could lead to spores delaying their ejection to avoid being sacrificed, simulations and asymptotic analysis show that, close the fruit body, ejected spores form a sheet-like jet that advances across the fruit body as more spores are ejected. By ejecting on the arrival of the sheet spores maximize \emph{both} their range and their contribution to the cooperative wind.
\end{abstract} 
 
 
\section{Clips}
The accompanying fluid dynamics video is available in (reduced) mpeg-1 and (full-size) mpeg-2 versions from the \href{http://hdl.handle.net/1813/14102}{eCommons} repository. The video includes 3 experimental and 3 numerical clips:
 
\begin{enumerate} 
\item Experimental video of a spore jet created by synchronized ejection from \emph{Sclerotinia sclerotiorum} fruit bodies (0.4$\times$ natural speed). For scale the petri dish containing the fruit bodies has a diameter of 10 cm.
\item Experimental video of \emph{S. sclerotiorum} spore jet impacting upon a glass slide (0.5$\times$ natural speed).
\item Direct numerical simulation of the spore jet produced by a fruit body of diameter 4 mm (natural speed). Surfaces show the azimuthal component of the vorticity. For scale the maximum length of jet is 95 mm.
\item Direct numerical simulation of spore trajectories (0.3$\times$ natural speed). Spores are color-coded according to time of ejection: red spores are ejected first, and green spores are ejected last. Only green (last ejected) spores enjoy maximum range enhancement from the cooperative behavior. Note different scalings of the horizontal and vertical axes.
\item High speed video of spore ejection from \emph{Ascobolus} cf. \emph{fulfarescens} (0.03$\times$ natural speed). For scale horizontal dimension is approximately 2 mm.
\item Direct numerical simulation of spore ejection from an 1 mm fruiting body (0.5$\times$ natural speed). Color-coding shows vorticity of the air flow created by the spores. Note different scalings on horizontal and vertical axes. 
\end{enumerate} 
\end{document}